\documentclass[%
 reprint,
 amsmath,amssymb,
 aps,
]{revtex4-2}
\usepackage{quantikz}
\usepackage[T1]{fontenc}
\usepackage[utf8]{inputenc}
\usepackage{cmap}
\usepackage{amsmath}
\usepackage{amssymb}
\usepackage{amstext}
\usepackage{amsfonts}
\usepackage{mathtools}
\usepackage{hyperref}
\usepackage{todonotes}
\usepackage{physics}
\usepackage{amsthm}
\usepackage{enumitem}
\usepackage{graphicx}
\usepackage{dcolumn}
\usepackage{bm}
\usepackage{hyperref}
\usepackage[capitalise]{cleveref}
\hypersetup{
 colorlinks=true,
 linkcolor=blue,
 citecolor=blue,
 filecolor=blue,
 urlcolor=blue,
 anchorcolor=blue,
 bookmarksopen=true,
 bookmarksopenlevel=1,
 pdfpagemode=UseOutlines,
 pdfstartview={XYZ null null 1},
 linktocpage=true,
}
\usepackage{tikz}
\usetikzlibrary{quantikz}
\usepackage{graphicx}
\usepackage{dcolumn}
\usepackage{bm}
\usepackage{xcolor}

\newcommand{\bigO}[1]{O(#1)}

\begin{document}
\title{Quantum computing for data analysis in high energy physics}
\thanks{This manuscript has been authored by UT-Battelle, LLC under Contract No.~DE-AC05-00OR22725 with the U.S. Department of Energy. The United States Government retains and the publisher, by accepting the article for publication, acknowledges that the United States Government retains a non-exclusive, paid-up, irrevocable, world-wide license to publish or reproduce the published form of this manuscript, or allow others to do so, for United States Government purposes. The Department of Energy will provide public access to these results of federally sponsored research in accordance with the DOE Public Access Plan (\url{http://energy.gov/downloads/doe-public-access-plan}).}%

\author{Andrea Delgado}
\email{delgadoa@ornl.gov}
\author{Kathleen E. Hamilton}
\author{Prasanna Date}
\affiliation{
 Oak Ridge National Laboratory, Oak Ridge, Tennessee USA}
\author{Jean-Roch Vlimant}
\email{jvlimant@caltech.edu}
\affiliation{California Institute of Technology}
\author{Duarte Magano}
\affiliation{Instituto Superior Técnico, Universidade de Lisboa, Portugal}
\affiliation{Instituto de Telecomunicações, Physics of Information and Quantum Technologies Group, Lisbon, Portugal}
\author{Yasser Omar}
\affiliation{Instituto Superior Técnico, Universidade de Lisboa, Portugal}
\affiliation{Instituto de Telecomunicações, Physics of Information and Quantum Technologies Group, Lisbon, Portugal}
\affiliation{Portuguese Quantum Institute, Portugal}
\author{Pedrame Bargassa}
\affiliation{Portuguese Quantum Institute, Portugal}
\affiliation{Laboratorio de Instrumentaçāo e Fisica Experimental de Particulas, Lisbon, Portugal}
\author{Alessio Gianelle}
\affiliation{INFN Sezione di Padova, Padova, Italy}
\author{Lorenzo Sestini}
\affiliation{INFN Sezione di Padova, Padova, Italy}
\author{Donatella Lucchesi}
\affiliation{Università degli Studi di Padova, Padova, Italy}
\affiliation{INFN Sezione di Padova, Padova, Italy}
\author{Davide Zuliani}
\affiliation{Università degli Studi di Padova, Padova, Italy}
\affiliation{INFN Sezione di Padova, Padova, Italy}
\affiliation{European Organization for Nuclear Research (CERN), Geneva, Switzerland}
\author{Davide Nicotra}
\affiliation{Universiteit Maastricht, Maastricht, Netherlands}
\author{Jacco de Vries}
\affiliation{Universiteit Maastricht, Maastricht, Netherlands}
\author{Domenica Dibenedetto}
\affiliation{Universiteit Maastricht, Maastricht, Netherlands}
\author{Miriam Lucio Martinez}
\affiliation{Universiteit Maastricht, Maastricht, Netherlands}
\author{Eduardo Rodrigues}
\affiliation{University of Liverpool, Liverpool, United Kingdom}
\author{Carlos Vazquez Sierra}
\affiliation{European Organization for Nuclear Research (CERN), Geneva, Switzerland}
\author{Anthony Francis}
\affiliation{Institute of Physics, National Yang Ming Chiao Tung University, Hsinchu 30010, Taiwan}
\affiliation{Theoretical Physics Department, CERN, CH-1211 Geneva 23, Switzerland}
\author{Jesse Thaler}
\affiliation{Center for Theoretical Physics, Massachussetts Institute of Technology, Cambridge, MA}
\author{Carlos Bravo-Prieto}
\affiliation{Quantum Research Centre, Technology Innovation Institute, Abu Dhabi, UAE}
\affiliation{Department de Fisica Quantica i Astrofisica and Institut de Ciencies del Cosmos (ICCUB), Universitat de Barcelona, Barcelona, Spain}
\author{Oriel Kiss}
\affiliation{European Organization for Nuclear Research (CERN), Geneva, Switzerland}
\author{Sofia Vallecorsa}
\affiliation{European Organization for Nuclear Research (CERN), Geneva, Switzerland}

\author{Su Yeon Chang}
\affiliation{European Organization for Nuclear Research (CERN), Geneva, Switzerland}
\affiliation{Institute of Physics, Ecole Polytechnique F\'ed\'erale de Lausanne (EPFL), Lausanne, Switzerland}

\author{Jeffrey Lazar}
\affiliation{Department of Physics \& Laboratory for Particle Physics and Cosmology, Harvard University, Cambridge, MA}
\affiliation{Department of Physic, University of Wisconsin-Madison, Madison, WI}

\author{Carlos A. Arg\"{u}elles}
\affiliation{Department of Physics \& Laboratory for Particle Physics and Cosmology, Harvard University, Cambridge, MA}

\author{Jorge J. Martínez de Lejarza, Leandro Cieri, Germán Rodrigo}
\affiliation{
  Instituto de F\'{\i}sica Corpuscular, Universitat de Val\`encia - Consejo Superior de Investigaciones Cient\'{\i}ficas, Parc Cient\'{\i}fic, E-46980 Paterna, Valencia, Spain}

\date{\today}

\begin{abstract}
Some of the biggest achievements of the modern era of particle physics, such as the discovery of the Higgs boson, have been made possible by the tremendous effort in building and operating large-scale experiments like the Large Hadron Collider or the Tevatron.
In these facilities, the ultimate theory to describe matter at the most fundamental level is constantly probed and verified.
These experiments often produce large amounts of data that require storing, processing, and analysis techniques that often push the limits of traditional information processing schemes.
Thus, the High-Energy Physics (HEP) field has benefited from advancements in information processing and the development of algorithms and tools for large datasets.
More recently, quantum computing applications have been investigated in an effort to understand how the community can benefit from the advantages of quantum information science.
In this manuscript, we provide an overview of the state-of-the-art applications of quantum computing to data analysis in HEP, discuss the challenges and opportunities in integrating these novel analysis techniques into a day-to-day analysis workflow, and whether there is potential for a quantum advantage.
\end{abstract}

\maketitle
\newpage

\tableofcontents

\section{\label{sec:introduction} Introduction}
Particle physics has the ambitious goal of uncovering the most fundamental constituents of the Universe and deciphering the rules that mediate their interactions.
Vast and complex accelerators are being developed to elucidate the dynamical basis of these fundamental constituents of matter.
At these large-scale facilities, high-performance data storage and processing systems are needed to store, access, retrieve, distribute, and process experimental data.
Experiments like the ones at the Large Hadron Collider (LHC) are incredibly complex, involving thousands of detector elements that produce raw experimental data at rates over a Tb/sec, resulting in the annual production of datasets in the scale of hundreds of Terabytes to Petabytes.
Beyond collider physics, a significant amount of data is also expected from upcoming neutrino experiments, \textit{e.g.} DUNE or IceCube-Gen2, and cosmological surveys, \textit{e.g.} DESI.
In addition, manipulating these complex datasets into summaries suitable for the extraction of physics parameters and model comparison is a time-consuming and challenging task. 

The high-energy physics (HEP) community has a long history of working with large datasets and applying advanced statistical techniques to analyze experimental data in the energy, intensity, and cosmic frontiers.
With the ever-increasing volume of data generated by HEP experiments, the community needs a significant breakthrough in the information processing systems to continue its successful journey into understanding the fundamental components of our universe.
Tools developed in quantum information science (QIS) could provide a viable solution.
Recently, alternative methods for detector simulation and data analysis tasks have been explored, like machine learning applications and QIS.
QIS is a rapidly developing field focused on understanding information analysis, processing, and transmission using quantum mechanical principles and computational techniques.
QIS can address the conventional computing gap associated with HEP-related problems, specifically those computational tasks that challenge CPUs and GPUs, such as efficient and accurate classification and simulation schemes.
In addition, quantum computing offers unique advantages over classical computing in machine learning and optimization.
Nonetheless, adapting these new technologies to the analysis of HEP data requires developing domain-specific tools and algorithms, such as quantum machine learning (QML) algorithms tailored to HEP applications. In References \cite{Guan_2021, Wu2022}, a review focusing on QML applications to HEP data analysis is presented. Nonetheless, the scope of this manuscript is broader, to include optimization algorithms.

In the following, we discuss the status and prospects for quantum computing for data analysis in HEP.
The emphasis is on prospects, starting from a thorough review of the current results and the related literature, thus illustrating the current status in each of the presented categories.
The aim is to underline the challenges faced and to highlight possible directions for future studies.
The manuscript is divided into sections covering a particular application and diving into the algorithms and models used.
We start with a short introduction to the most common physical realization of quantum computing in Section~\ref{sec:quantum_computing}.
The subsequent sections deal with applications of quantum computing to typical HEP tasks such as object reconstruction (Section~\ref{sec:obj_reco}), classification (Section~\ref{sec:classification}), and data generation or augmentation (Section~\ref{sec:data_generation}). 
In Section~\ref{sec:quant-insp}, we present a discussion of quantum-inspired algorithms.
In Section~\ref{sec:challenges}, we provide a comprehensive list including the challenges encountered in the application of quantum computing to classical data from HEP experiments.
Finally, in Section~\ref{sec:outlook}, we present our views on the future of advancing HEP with quantum computing.

\section{\label{sec:quantum_computing} Quantum Computing}
Quantum computing involves using the quantum mechanical properties of matter, such as entanglement and superposition, to process information.
Furthermore, it is based on the \emph{qubit}, or quantum bit, instead of the classical two-state binary bits.
Unlike bits, qubits can be in a superposition of both states, ``1'' and ``0'',  leading to an exponential increase in the amount of information encoded into these fundamental information processing units.
For this reason, the information processing capacity of quantum computing is significantly superior to traditional computational technologies, allowing (in theory) for the reduction in execution time for specific computing tasks.
Current-day quantum processors are prone to noise and small in scale, limiting the number of applications these devices can be used.
In what follows, a short description of the leading quantum technologies available to domain scientists is presented.

\subsection{Quantum Circuit Model}
In the early 1980s, Richard Feynmann introduced the concept of quantum computing  as a universal quantum simulator~\cite{Feynman1982}.
The materialization of this idea did not happen until 1985 when the concept of a \textit{quantum Turing machine} was coined by David Deutsch~\cite{1985Deutsch}.
This formal definition of quantum computing is based on what we know today as a \textit{universal quantum computer}, based on qubits and unitary transformations or quantum gates. 
In this model, we begin in an initial ``zero'' state, apply a sequence of quantum gates chosen from a set of allowed gates, each acting on one or two qubits at a time, and finally output the outcome of a measurement in the computational basis.

\subsection{Quantum Annealing}
%
%


Another paradigm of quantum computing is quantum annealing (QA)~\cite{Kadowaki1998, farhi2000quantum}.
QA is a technique to solve combinatorial optimization problems by encoding the solution as the ground state of some Hamiltonian. 
The ground state solution is reached by initializing the quantum system in the ground state of another known, and easy to control, Hamiltonian. 
We then let the system evolve by slowly changing the Hamiltonian to the target one. 
The quantum adiabatic theorem~\cite{Ambainis2006} guarantees that there is a minimum evolution time for which the final state is as close to the solution state as we desire; this time depends on the smallest energy gap attained between the ground state and the first excited state of the slowly changing Hamiltonian.

\begin{figure}
    \centering
    \includegraphics[width=0.85\linewidth]{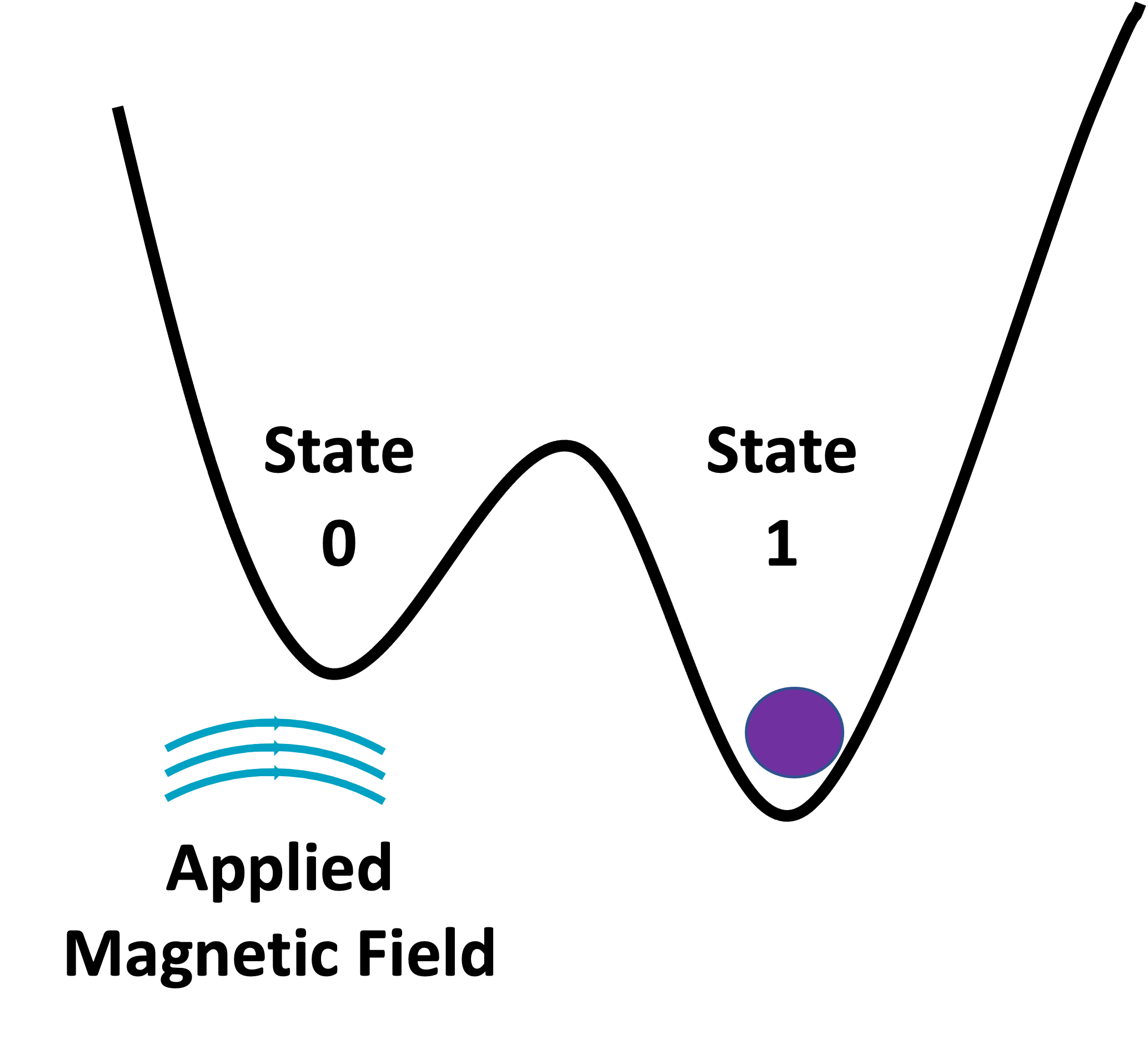}
    \caption{Schematic diagram of the effect of applying a magnetic field to a system initially prepared in superposition.
    After the magnetic field is modified, the energy diagram of the single-qubit system is tilted towards state ``1'', resulting into a higher probability of the system collapsing to state ``1'' after a measurement.}
    \label{fig:qa}
\end{figure}

\subsection{Continuous-Variable Quantum Computing}

Most of the quantum computing applications to data analysis in HEP reviewed in this manuscript focus on implementations in quantum annealers or gate-based models.
While these models are very successful, another scheme explored much less than the other two QC paradigms is the continuous-variable (CV) model~\cite{PhysRevLett.82.1784}.
CV differs from its use of \textit{qumodes} over qubits.
The qumodes constitute an infinite-dimensional object instead of discrete qubits and constitute a more natural choice to simulate bosonic and continuous systems.
Programming photonic quantum devices proceed in a very similar way to qubit-based devices, with both allowing the construction of circuits from quantum gates. 

\section{Object Reconstruction}
\label{sec:obj_reco}
As mentioned in Section~\ref{sec:introduction}, experimental HEP deals with the analysis of large amounts of experimental data produced at HEP experiments.
One of the main tasks in any HEP data analysis workflow is the construction of \emph{physics objects} amenable to analysis from the signals produced in a particle detector - \textit{i.e.}, how to translate the raw detector data into an object whose properties reflect the kinematics of the initial interaction (a $p-p$ or a heavy-ion collision, or a neutrino event). 

\subsection{Tracking}

Hadron colliders, such as the LHC, accelerate counter-rotating beams of hadrons in tightly packed bunches.
These hadron bunches cross at designated interaction regions surrounded by tracking detectors.
At each bunch crossing, several hadron-hadron collisions take place along the direction of the beam axis.
Every hadron collision produces a myriad of secondary particles scattered in all directions.
As the produced charged particles cross the detector's multiple layers of sensors, they leave signals of their passage, known as \emph{hits}.
The collection of the hits left by a particle is called that particle's \emph{track}.
The \emph{tracking} problem is to recover the tracks from a set of hits.
It constitutes a key task in the analysis of accelerator experiments, essential to studying the underlying physics.

At the LHC, the proton bunches cross with an interval of twenty-five nanoseconds~\cite{JINST}.
In the conditions of the Run2 of the LHC, there are around thirty-five proton-proton collisions at each bunch crossing, and each collision may produce a few thousand hits.
As a result, vast amounts of data are generated, making its analysis one of the most computationally demanding activities in experimental HEP ~\cite{Albrecht_Roadmap}.
Moreover, this demand is expected to grow dramatically after 2026 with the upcoming high-luminosity phase of the LHC~\cite{HL-LHC} (HC-LHC), when we will typically have around two hundred proton-proton collisions per bunch crossing, and even more so in future accelerator machines, such as the Future Circular Collider~\cite{FCC}.

Therefore, it is crucial to develop efficient tracking methods, possibly requiring completely novel technological paradigms.
In this section, we review the existing proposals for track reconstructions based on quantum computing.

\subsubsection{Tracking with Amplitude Amplification}

The Combinatorial Kalman Filter method is at the base of several reconstruction programs in high-energy physics~\cite{JINST, ATLAS, Braun2018}.
Ref.~\cite{magano2021quantum} analyzed this technique from a computational complexity perspective, identifying four fundamental routines: seeding, track building, cleaning, and selection.  
The seeding stage forms initial rudimentary track candidates, known as seeds, with just a few hits.
Then, the track building stage extrapolates the seeds' trajectories along the expected path of the particles, forming track candidates by adding compatible hits from successive detector layers.
To avoid having multiple tracks describing the same particle, a cleaning process is applied to remove the ones that are too similar.
Finally, we select only the tracks that respect some quality criteria based on the quality of the fit between the trajectory and the corresponding hits.

Ref.~\cite{magano2021quantum} proposes a quantum implementation of the Combinatorial Kalman Filter, relying on a fundamental quantum routine known as amplitude amplification~\cite{brassard2002quantum} (a generalization of Grover's algorithm~\cite{Grover1996}), which allows for polynomial speedups for certain unstructured search problems.
The general idea is that we can use this technique to find the seeds or hits that satisfy specific properties faster than what would be possible with classical brute-force search.
For both the seeding and track building stages, it is possible to reproduce the same output as the Combinatorial Track Finder (up to bounded error probability) with lower quantum computational complexity.
Ref.~\cite{magano2021quantum} further shows that if we do not register the outputs of the individual stages, but are only interested in the final reconstructed tracks, then an even stronger quantum advantage can be reached.

Ref.~\cite{magano2021quantum} provides a rigorous proof of quantum speedup for HEP data analysis.
However, its results refer to the asymptotic regime of infinitely-many hits, which may be of limited interest for real instances of track reconstruction.
Moreover, this proposal depends on the availability of a quantum random access machine (QRAM)~\cite{QRAM} that allows coherent access to the hits' data.
Although there have been proposals of physical architectures for QRAM~\cite{QRAM_architectures}, there are still significant challenges to overcome before such a device can be realized in practice.
Finally, the reached speedups have been found to be mild; quantitatively, $O(n^4)$ classically versus $O(n^3)$ quantumly under certain assumptions, where $n$ denotes the number of recorded hits.

As the algorithms proposed in~\cite{magano2021quantum} require fault-tolerant computing, it has not been possible to test them in real quantum hardware.
Therefore, it remains unclear at which scales the complexity advantage may actually yield an advantage in processing time.
In general, fault-tolerance and QRAMs lie beyond the NISQ era, so it is hard to estimate when this type of strategy might have a practical impact in HEP.

\begin{figure}
    \centering
    \includegraphics[width=0.85\linewidth]{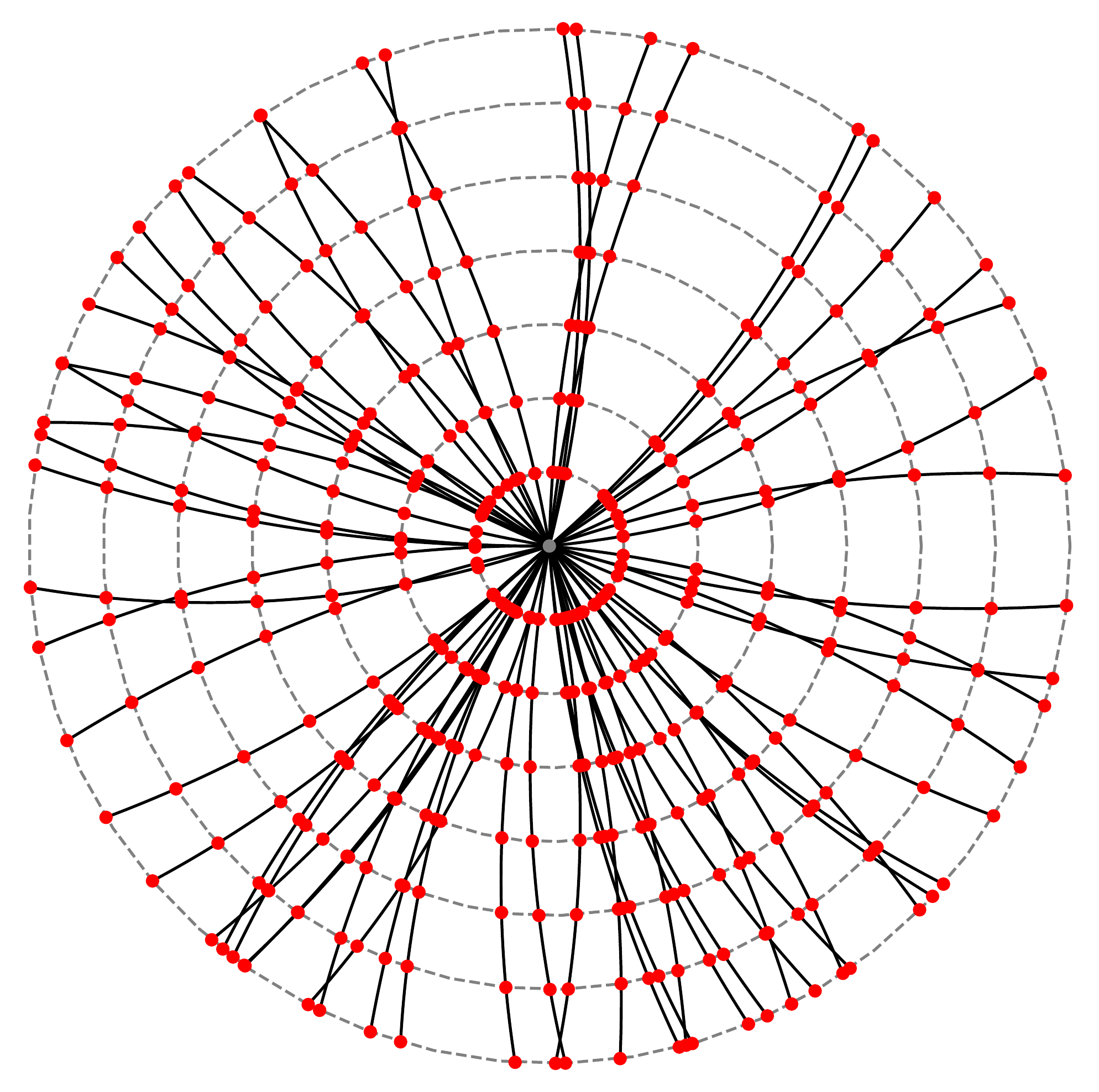}
    \caption{Illustration of the tracking problem.
    Transverse view of a tracking detector with cylindrical  layers  (dashed  grey  lines).
    The input to tracking is a set of hits (red circles) corresponding to detections of the particles' passage.
    We recover the original trajectories (black lines) by grouping hits that belong to the same particle, i.e., by reconstructing the particles' tracks.}
    \label{fig:qa}
\end{figure}

\subsubsection{Tracking with Quantum Annealing}

The work of Ref.~\cite{magano2021quantum} consists of a direct ``quantization'' of a classical algorithm (the Combinatorial Kalman Filter), which might not be the best path to establishing a significant advantage in quantum computing for HEP problems.
Instead, a more successful approach could be obtained by breaking the direct correspondence with the classical setting and designing completely new tracking algorithms that inherently take advantage of the features of quantum processors.
Some of the first attempts in this direction~\cite{Zlokapa2021,Bapst2019,das2020track, Quiroz2021} were conceived within the quantum annealing model.

In particular, Ref.~\cite{Zlokapa2021} and Ref.~\cite{Bapst2019} have formulated track reconstruction as a quadratic unconstrained binary optimization (QUBO) problem, which can be naturally mapped to a quantum annealer.
In Ref.~\cite{Zlokapa2021}, the QUBO variables correspond to all possible edges between hits.
The idea is that, in the optimal solution, the variables assigned with $+1$ connect hits that are left by the same particle.
For this purpose, Ref.~\cite{Zlokapa2021} adapts the energy function of the Denby-Peterson network method~\cite{denby88,Peterson89} to HL-LHC configuration.
In contrast, in Ref.~\cite{Bapst2019} the binary variables correspond to hit triplets and introduce a compatibility measure between hit triplets to define the QUBO problem.
The use of hit triplets instead of doublets comes with a greater classical pre-processing cost while possibly increasing the performance of the reconstruction.

With both approaches, there are polynomially many coupling coefficients that need to be classically pre-processed to form the QUBO Hamiltonian -- $O(n^3)$ for Ref. ~\cite{Zlokapa2021} and $O(n^5)$ for Ref.~\cite{Bapst2019}, where $n$ is the number of hits.
Moreover, the annealing time is likely to scale exponentially with respect to the number of hits.
Hence, it remains unclear whether these methods could exhibit quantum speedup over the classical counterpart (\textit{i.e.}, simulated annealing).
Nevertheless, the simulations done at the D-Wave's machines~\cite{DWave19} show competitive efficiency and purity for some toy datasets.

Ref.~\cite{das2020track} addresses the problem of track clustering, not track reconstruction \textit{per se}.
This is the first step in reconstructing the positions of hadronic interactions, also known as primary vertices.
The algorithm takes as input a set of positions $z_i$, where $i$ is the track index, along the beam axis corresponding to the points where the reconstructed tracks approach it most closely.
The goal is to associate each track with an element of a list of possible primary vertices.
In the algorithm of Ref.~\cite{das2020track}, the annealing variables are the entries of a binary association matrix between the tracks and the candidate vertices, and the annealing Hamiltonian penalizes assignments of tracks $i$ and $j$ to the same vertex if $ \vert z_i -z_j \vert$ is large.
In total, one needs to pre-compute $O(n_v n_t (n_v + n_t))$ coupling coefficients, where $n_v$ is the number of candidate vertices and $n_t$ is the number of tracks. As in Ref.~\cite{Zlokapa2021,Bapst2019}, there is no guarantee on scaling the required annealing schedule.
Ref.~\cite{das2020track} tests their algorithm with small datasets on a D-Wave's quantum computer, reporting clustering precisions close to the results obtained with simulated annealing.

In contrast, Ref.~\cite{Quiroz2021} studies the problem of track classification, that is, the problem of distinguishing between signal and background events from the hits' data.
This seeks to accelerate track reconstruction by isolating the signals of interest.
For this purpose, Ref.~\cite{Quiroz2021} utilizes both associative memory models and content addressable memory models in conjunction with quantum annealing.
In this setting, a collision event is characterized by a detection pattern -- the detector is divided into discrete positional segments and we record the number of hits in segment.
Given a history of detection patterns, the QUBO coefficients are assigned based on a specified learning rule.
The computational complexity of training the model scales as $O(N^3 p^2)$, where $N$ is the length of the patterns and $p$ is the number of training patterns.
Again, it is difficult to estimate how the annealing time scales with the model parameters.
The performance of the memory recall is dependent on $N$ and $p$, as well as on the noise levels and the efficiency of the detections.
Running simulations on D-Wave's quantum hardware, Ref.~\cite{Quiroz2021} report that, for small problem instances, these models can reach accurate classification results.

In summary, none of these proposals based on quantum annealing manage to conclusively establish a quantum advantage for track reconstruction over the classical counterparts.
However, the simulations on currently available quantum hardware, particularly on D-Wave's annealing machines, already reveal competitive results, as long as one is restricted to small problem sizes.
A complete characterization of the performance and scaling of these annealing-based solutions will require larger programmable quantum annealers.

\subsubsection{Tracking with Neural Networks}

Ref.~\cite{Tysz2021} seeks to leverage the sparse nature of the tracking data with hybrid quantum-classical neural networks.
As a pre-processing step, they generate a graph from the detector's data, where the particle's hits become the nodes, and the possible track segments between hits become the links.
Then, a neural network model takes this graph as input and gives as output the probability of each edge linking two consecutive hits left by the same particle.
This neural network combines both classical and quantum layers, the latter realized in the form of parametrized quantum circuits.

Evidently, the reconstruction performance depends on the number of hidden classical $n_C$ and quantum $n_Q$ dimensions, and on the number of training iterations $n_I$.
For small values of $n_C$, $n_Q$, and $n_I$, Ref.~\cite{Tysz2021} shows that the hybrid networks closely match the results of the classical model, at least for the simplified datasets that were considered.
Given the current limitations in quantum hardware, it is difficult to assess whether increasing the size of the network ($n_C$, $n_Q$, and $n_I$) would significantly improve the results.
It also remains unclear whether neural network algorithms will ever achieve the precision of the Combinatorial Kalman Filter methods.

Finally, the complexity of the algorithm depends on the network parameters, and as such these, become hyperparameters in the optimization scheme. A further discussion on this topic can be found in Section \ref{sec:vqc}.

\subsection{Jets}
In collider particle physics, a jet is a collection of particles collimated into a roughly cone-shaped region.
Jets arise from the fragmentation of quarks and gluons produced in high-energy collisions.
During the collision, the QCD confinement the quarks and gluons are subjected to, is broken, yielding a spray of color-neutral particles that can be experimentally measured in particle detectors.
Jet clustering algorithms are employed to estimate the kinematics of the particle that initiated the jet.
These clustering schemes combine the observed particles into a collective jet object for further study.
In electron-positron collisions, the dominant event topology involves two back-to-back jets resulting from the fragmentation of a quark and an anti-quark.
This motivates the partitioning of the event particles into two hemisphere jets, which can be accomplished using event shapes like \textit{thrust}~\cite{BRANDT196457,PhysRevLett.39.1587}.
The calculation of thrust can be very expensive computationally, scaling like $O(N^3)$~\cite{YAMAMOTO1983597} for an event with $N$ particles, though using a method introduced in Ref.~\cite{Salam_2007}, it is possible to improve this to $O(N^2 \log N)$.
On a universal quantum computer, thrust can be computed in $O(N^2)$~\cite{PhysRevD.101.094015} using a strategy based on Grover search.
Alternatively, Ref.~\cite{PhysRevD.101.094015} showed how thrust can be phrased as a QUBO problem suitable for quantum annealing.

\begin{figure}
    \centering
    \includegraphics[width=0.65\linewidth]{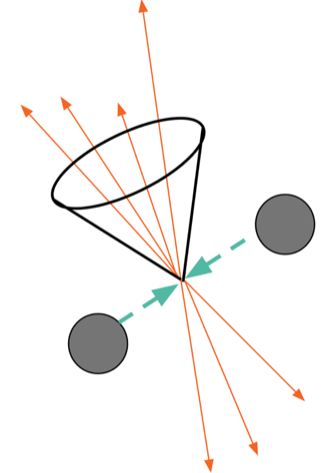}
    \caption{Schematic diagram of a particle jet. A collision between two highly energetic particles (gray circles) results in a spray of color neutral hadrons (orange arrows) collimated into a roughly cone-shaped region.}
    \label{fig:jet}
\end{figure}

In Ref.~\cite{DelgadoThaler2022}, the thrust-based quantum annealing for jet clustering is benchmarked on the D-Wave Advantage 1.1 QPU.
Algorithmic improvements like reverse annealing showed promise on this problem, but the most significant gains came from tuning the annealing parameters such as annealing time, the number of anneals, and relative chain strength.
The QPU performance was also compared to hybrid and classical solving strategies, with simulated annealing always performing the best.
Nonetheless, for event sizes smaller than 22 particles, the classical and quantum annealing solvers display a similar performance.
These results hint at the limitations of current quantum annealing devices in terms of connectivity.
Problems involving many spin variables and all-to-all connectivity, like the thrust problem, perform poorly on this device.

Based on the quantum annealing-based algorithm proposed in~\cite{PhysRevD.101.094015}, an extension to the QUBO formulation for thrust calculation is presented in Ref.~\cite{pires2020adiabatic}.
This extension is based on the angular distance between two particles in a given event, penalizing the assignment of two particles located on the same hemisphere of the true partition.
This approach displays a better performance when tested on the D-Wave's QPU and when compared to the original formulation in Ref.~\cite{PhysRevD.101.094015}.
It should be noted that these benchmark studies were limited to a low number of annealing runs due to limited access to the QPU.

On the other hand, algorithms based on digital quantum computing have also been proposed. The algorithms in Refs.~\cite{pires2021digital, PhysRevD.101.094015} are not suitable for implementation on NISQ devices due to the need for a QRAM architecture to access the information about the particles on the event for further processing. Another promising study~\cite{deLejarza:2022bwc} deals with the quantum version of three clustering algorithms found in the classical literature: \textit{k-means} \cite{MacQueen1967}, \textit{affinity propagation}~\cite{Brendan2007}, and the \textit{$k_T$} jet clustering algorithm \cite{PhysRevD.48.3160}. Two quantum subroutines are introduced: the first computes the Minkowski distance between particles, and the second tracks the maximum in a long-tailed distribution. These algorithms were applied to simulated data for a typical LHC collision setting and obtained efficiencies comparable to their classical counterparts. In particular, the quantum-\textit{$k_T$} version is a conceptually more straightforward algorithm with a similar execution time when compared to subroutines in the FastJet library. 

As a summary for this section, track reconstruction and jet clustering tasks in HEP deal with the clustering of detector information into higher-level objects that contain information about the interaction that triggered the detector response.
The main challenge is the number of elements to be clustered.
Quantum computing applications to the problems of track reconstruction and jet clustering are limited by the size and connectivity of currently available devices.
Notably, these tasks require either a quantum annealer with all-to-all connectivity or a QRAM-like protocol for accessing the information of individual elements in the clustering sample in coherent superposition.

\section{Classification}
\label{sec:classification}
Once detector data has been processed and higher-level objects have been constructed, the next step in a HEP analysis is using statistical techniques to extract the signal of interest and suppress background as much as possible. Thus, efficient classification tools are crucial in HEP data analyses. Classification algorithms are not limited to the discrimination events of interest (signal) from other processes (background). In addition, reconstructed particles also undergo a process of classification to be assigned a label or identification according to their type and kinematic properties. Jets need to be classified according to their origin. 

Classification tasks are usually addressed with Multivariate Analysis Techniques (MVA).
Current physics analyses employ analytic methods like the Matrix-Element Method~\cite{Artoisenet2013}, as well as Boosted Decision Trees (BDTs) and Neural Networks (NN).
In this section, new approaches to classification tasks based on different quantum computing architectures and algorithms are reviewed, which may yield better performance than their classical counterparts.

\subsection{Quantum annealing applications}
\label{sec:anneal}

One of the main and first quantum computing models used for classification in HEP is quantum annealing.
In this binary approach, each discriminating variable $x_i$ is transformed into a weak classifier $c_i$, and has an associated spin $s_i$.
A strong classifier is then built, which is the linear combination of all weak classifiers and spins.
The classification is then expressed in terms of the optimal set of spins minimizing the energy of a fully connected Ising Hamiltonian.
This set of spins is found after an iterative annealing process where qubits are strongly coupled in a chain to represent a spin $s_i$.
Correlations among the discriminating variables are possible via the $c_i \cdot c_j$ coupling terms in the Hamiltonian, and are rendered further present via operations among them, which are then transformed in weak classifiers.
Fixing variable scheme and a cutoff on the coupling terms are used to reduce the size of the Ising model to be encoded on the annealer.
This approach is followed in Ref.~\cite{Mott2017} where the signal from Higgs boson decays is separated from the background.
In this work, quantum annealing achieves a better result for a small number of training events but is outperformed by classical machine learning (ML) tools for a large number of events.
The basic approach of quantum annealing for classification \cite{Mott2017} is further enhanced in Ref.~\cite{qamlz}, where the procedures of \textit{zooming} and \textit{augmentation} are applied.
While the zooming shifts and narrows the region of search in the space of spins, the augmentation multiplies the number of binary weak classifiers $c_i$ based on the shape of each variable $x_i$, effectively providing better discrimination.
These two steps significantly improve the result of the Higgs classification problem; however, they are still outperformed by classical ML approaches.
The principle of zoomed and augmented annealing is applied to the new classification problem of supersymmetric top quark versus SM events~\cite{Bargassa2021}.
Here, the choice of the discriminating variables is based on a metric incorporating the full statistical and systematic uncertainties of a counting experiment in HEP.
This results in a relatively reduced (only 17) but well-performing set of discriminating variables.
Different such sets of variables are tested with different augmentation schemes.
Finally, in order to place the quantum-based classifier on a footing as equal as possible to the classical Boosted Decision Trees (BDT), the discriminating variables are decorrelated by being passed through a principal component analysis before being fed to the quantum annealer.
The study of Ref.~\cite{Bargassa2021} achieves a classification that is at least as good as the best-known classical ML tool, here the BDT. It has to be noted that this result is attained for a rather large number of events in the training sample ($5\cdot10^4$), which is the typical size of samples used in HEP, and for which the first quantum based classifications are outperformed by classical approaches.
The results of Ref.~\cite{Mott2017,qamlz,Bargassa2021} are all obtained with the Chimera graph of D-Wave.
If a chain is broken within this device, the measure of the qubit chain is performed through a majority vote, which can lead to the selection of non-optimal sets of spins, and therefore to a possible loss of discriminating information.
This is a limitation of these first-generation quantum annealers.
A larger number of couplers in future machines will render each chain more stable and less prone to be broken, which will, in turn, allow more effective use of the discriminating information.

Quantum annealing can also be used to identify the topology of a signal event, where there is no hypothesis about the latter and where the mass of the new particles is inferred from their decay products.
Ref.~\cite{kim2021} looks at cases where two new objects would be produced at an LHC collision, with each decaying in a number of known particles.
The problem to solve is combinatorial, where the correct association of two groups of observed particles has to be made to reconstruct the invariant mass of the new particles.
The spin $s_i$ indicates whether a particle $i$ is the decay product of one or the other new particle.
The kinematic constraints of the new particles are formulated in the Ising Hamiltonian, which is encoded on a Pegasus graph of D-Wave, and whose energy is minimized.
The energy landscape of such a problem is complicated, and classical annealing will have a local minimum problem.
The combinatorial problem solved by quantum annealing reaches an accuracy higher than a classical algorithm for three different processes, thus indicating a quantum advantage.

The study in Ref.\cite{matchev2020} also uses an Ising Hamiltonian-based approach to perform a model-independent search for physics beyond the SM.
The kinematic space is partitioned in bins $i$, where the difference between the simulated SM prediction and observed data is measured, and where each spin $s_i$ can either be aligned (\textit{i.e.} of the same sign) or anti-aligned with the measured difference.
The Hamiltonian is expressed such that the ground state energy of the system is a measure of the goodness-of-fit.
Linear spin terms in the Hamiltonian are only sensitive to the difference mentioned above in each bin $i$.
On the other hand, $s_i \cdot s_j$ terms capture the interaction between neighboring bins, therefore being sensitive to spatial correlations between different kinematic regions, which in turn helps to differentiate between random noise and new physics signals.
This work relies on simulated annealing as a method for minimizing the energy of the Hamiltonian.
Toy experiments are generated where signal-plus-background and background hypotheses are tested.
The goodness-of-fit of different approaches is tested in terms of true- versus false-positive rates. Both with one- and two-dimensional toy experiments, the goodness-of-fit test statistic based on this quantum algorithm performs better than classical methods.
It has to be noted that this improved capacity to detect an anomaly versus an expectation is free on any assumption on the signal, and is thus model-independent.

The work presented in Ref.~\cite{caldeira2020restricted} explores a classification application of importance in cosmology, a galaxy morphology classification by training Restricted Boltzmann Machines (RBMs) in a quantum annealing device.
RBMs are, generally speaking, generative models and will be discussed in this context in Section~\ref{sec:quantum_gen_mods}.
In the classical setting, an RBM is a stochastic neural network that can learn a probability distribution over its set of inputs.
The study in Ref.~\cite{caldeira2020restricted} found that for small datasets and limited numbers in training repetitions, quantum annealing-based RBMs performed very well and outperformed the alternative classical algorithms studied, namely logistic regression and gradient boosted trees.
However, outside of these rather special training scenarios, RBMs (regardless of the classical or quantum nature of the training algorithm) did not outperform the gradient boosted tree algorithm.

\subsection{Variational quantum circuits}
\label{sec:vqc}

Variational Quantum Circuits (VQCs) are hybrid quantum-classical algorithms aiming to harness the strength and scalability of both computational paradigms.
In this architecture, classical computers are used for optimization and quantum computers for specific complicated tasks, such as calculating expectation values. 

A VQC can be viewed as a Quantum Neural Network (QNN) where the encoded quantum state goes through a circuit with different layers of gates that depend on parameters that will minimize a loss function through training.
In a VQC, once an initial state has been prepared through the encoding of classical data into a quantum state, a series of unitary transformations are applied through a circuit with layers that depend on trainable parameters $\vec{\theta}$ and act serially, mimicking the forward pass of a neural network, as first noted in~\cite{Abbas2021}.
As a classifier, a VQC can be trained from labeled data to classify new samples or, in a generative setting, to model correlations in the input data.
Recent studies have reported the application of variational architectures in the field of classification~\cite{Benedetti2019}, function approximation~\cite{Mitarai2018}, generative machine learning~\cite{dallaire2018quantum, zoufal2021variational}, metric learning~\cite{quantumembeddings}, deep-reinforcement learning~\cite{var_deeplearning} and sequential learning~\cite{qrnn}.

In the gate-based quantum computing model, variational algorithms are implemented using quantum circuits composed of a network of single and two-qubit operations, with rotation angles serving as variational parameters.
The implementation of a VQC usually takes place in three steps:

\begin{enumerate}
    \item First, an initial state is prepared using a feature map or unitary transformation $U_{\phi}(\vec{x})$ to encode the classical input data $\vec{x}$, into a quantum state.
    The input feature vectors become the rotational gates' arguments and remain fixed during the circuit evaluation.
    Some standard techniques include angle and amplitude encoding; see Ref.~\cite{jlazar2021} for more complex approaches.
    \item The second block of unitaries is a parameterized quantum circuit or variational form.
    It is given by $U(\vec{\theta})$ parameterized by gate angles $\theta$ and includes alternating layers of entangling and rotation gates.
    Each layer consists of a sub-circuit that depends on trainable parameters and an entanglement sub-circuit.
    The learnable parameters $\vec{\theta}$ will be optimized through gradient-based methods. For example, the commonly used gradient-based optimizer is Adam~\cite{adam_opt}.
    \item Then, a quantum measurement is performed on a subset (or all) of qubits to retrieve the information.
    If we run the circuit once and perform a single quantum measurement, it will yield a binary string, and it generally differs from what we will get if we prepare the circuit again and perform another quantum measurement due to the stochastic nature of quantum systems.
    However, if we prepare the same circuit and perform the quantum measurement several times, we will get the expectation values on each qubit.
    Furthermore, different bases can be used for the measurement.
    In most cases, the expectation value of the $\sigma_Z$ Pauli operator is used to obtain measurements on the computational basis.
\end{enumerate}

\begin{figure}
    \centering
    \includegraphics[width=\linewidth]{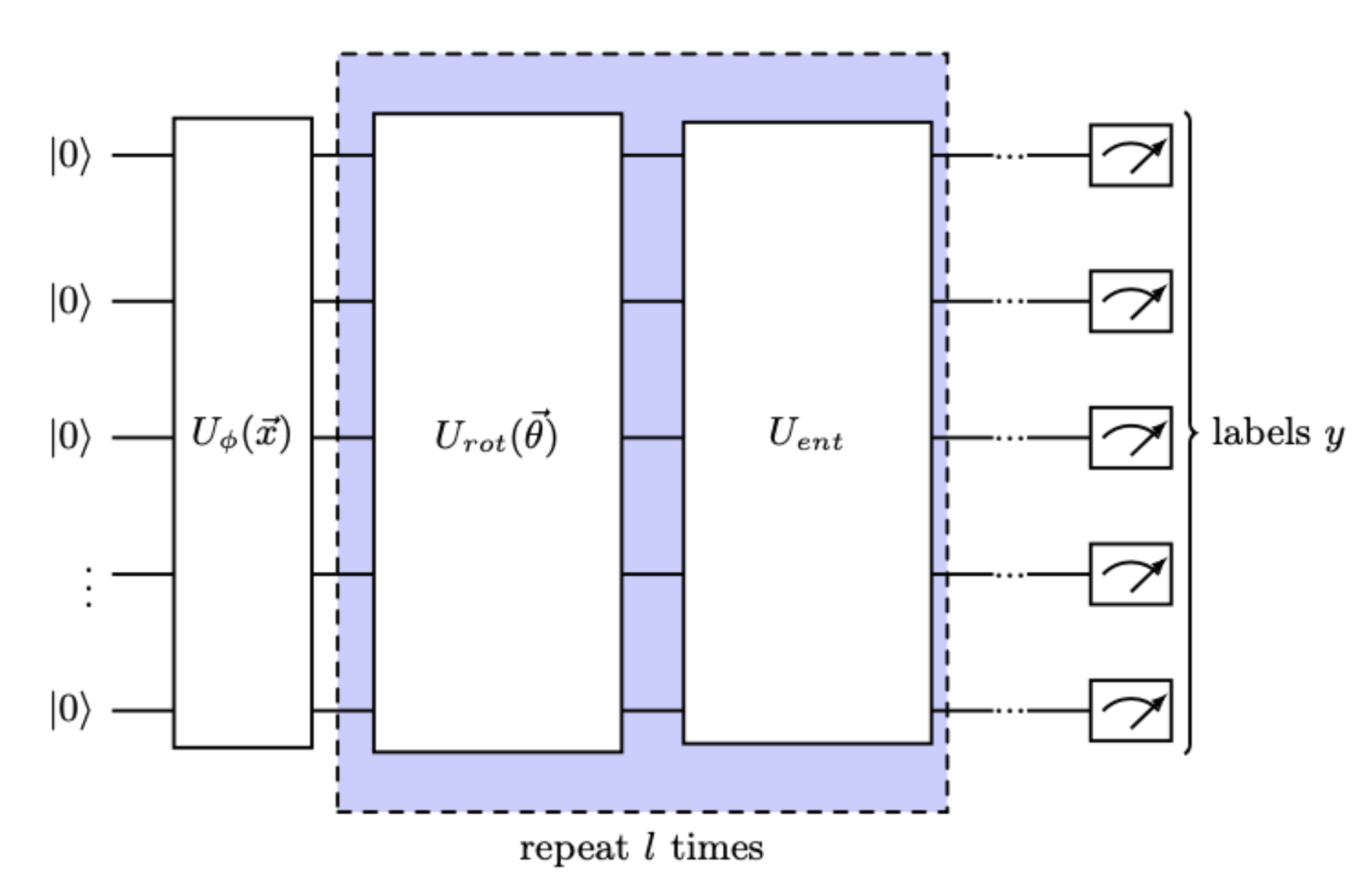}
    \caption{Schematic diagram of the main component in a Variational Quantum Circuit (VQC): (1) An encoding unitary $U_{\phi}(\vec{x})$, (2) a parameterized circuit with variational parameters with alternating layers of rotational ($U_{rot}(\vev{\theta})$) and entangling ($U_{ent}$) gates, repeated \emph{l} times, and, (3) a measurement and post-processing section.}
    \label{fig:vqc_circuit}
\end{figure}

In most cases, as we will see in some of the HEP applications discussed below, the VQC performance is evaluated by looking at the Receiver Operating Characteristics (ROC) curve and the Area Under the Curve (AUC). In what follows, applications of VQCs to the classification of HEP experimental datasets will be discussed.

The work in Ref.~\cite{Terashi2021} studies the classification of chargino-pair production via a Higgs boson versus the SM background.
In this study, two circuit designs or \emph{Ansatz} are tested to demonstrate the feasibility of QML for the event classification task in HEP data analysis.
The first choice is the circuit proposed in Ref.~\cite{Mitarai2018} and constructed using a time-evolution gate, denoted as $e^{-iHT}$, with the Hamiltonian $H$ of an Ising model with random coefficients and the series of $R_{X}$, $R_{Z}$ and $R_{X}$ gates, whose rotational parameters are updated during training.
This layout is repeated three times after applying a unitary that encodes the classical input data as rotational parameters in the $R_{Y}$ and $R_{Z}$ gates. The second choice of VQC is based on the circuit proposed in Ref.~\cite{Havlek2019}, which uses a layer of Hadamard and $R_{Z}$ gates to prepare the input state, followed by an entangling layer sandwiched between two layers of rotational $R_{Y}$ and $R_{Z}$ gates with trainable parameters. 
Measurement is performed on the first two and all the qubits using the Pauli Z operators for the first and second circuit designs, respectively.
The VQC is trained by minimizing the cross-entropy cost function as implemented in the \texttt{scikit-learn} package \cite{scikit-learn} and using the \texttt{COBYLA} optimization algorithm \cite{Powell1994}. 

The performance of the proposed VQCs is benchmarked against two classical ML algorithms: a BDT, as implemented on the Keras package \cite{chollet2015keras}, and a Deep Neural Network model based on a fully-connected feed-forward network composed of 2-6 hidden layers with 16-256 nodes each, implemented using the \texttt{Tensorflow} package \cite{tensorflow2015-whitepaper}.
The hyperparameters in the classical models are chosen to match the capabilities of the quantum models, and both methods  are applied to the same task and dataset size. It has to be noted that this classification is based on a small number of discriminating variables. The total number of internal parameters is 12, 20, and 28 for the 3-, 5-, and 7-variable classification, respectively. Concerning benchmarking on hardware, only the circuit design based on single and two-qubit gates is tested on a quantum backend. The VQC is run on the 20-qubit IBM Q Network quantum computer. It is noteworthy that this VQC, in the present stage, does not perform better with a larger number of features.

In Ref.~\cite{Belis2021}, a VQC approach is applied for the classification of $t\overline{t}H$ where a pair of top quarks are produced with a Higgs boson. 
In this study, two Auto-Encoder NNs (see Section~\ref{sec:generative_models}) are used to reduce the dimension of the feature space, and thus, ease the classical input feature encoding into quantum states.
This compressed representation is the input for the hybrid architectures introduced and the benchmark against non-quantum techniques. 

A \textit{data re-uploading} technique, introduced in Ref.~\cite{Schuld21_Fourier,perezsalinas:2020reuploading}, is used in an effort to reduce the number of quantum registers needed to encode the classical data into a quantum state.
In this approach, the quantum classification circuit comprises several repetitions of the traditional VQC scheme, each with its own classical inputs in the feature map and trainable parameters. 
The latter study is implemented in circuits comprising different number of qubits. Using the data re-uploading technique, the feature map  based on one and two-qubit gates is applied to load the classical variables.
Finally, the qubits are measured and used to classify the input.
Effectively, data re-uploading QNNs can be cast as partial Fourier series, where the size of the set of available frequencies grows linearly with the number of data encoding gates, while the coefficients are learned with the variational layers. In Ref. \cite{Kiss22_QNNFF}, a data re-uploading model is train to fit the potential energy surface and force field of small molecules. 

The study in Ref.~\cite{2021Blance} is a VQC approach to classify $t\overline{t}$ events versus the rest of the SM processes, using only two kinematic features.
An angle encoding scheme is used to prepare the two-qubit initial state.
The trainable block of the VQC consists of two layers of single rotation gates, followed by two CNOT gates, entangling the qubits in the circuit.
Finally, a Pauli $Z$ operator is applied on the first qubit, where the expectation value is taken.
The circuit is run repeatedly to obtain an estimate of the latter, where the size of the training sample is relatively small (1500 events).
Two different instantiations of the VQC are tested, where a classical and a quantum gradient descent are run and compared to the performance of NN with a gradient descent.
The quantum gradient optimization shows a faster convergence than the traditional gradient descent optimization and the classical neural network.
In terms of classification, the VQC is observed to perform better than the NN in high purity regimes, \textit{i.e.} when cutting on high values of the classifier's output.
The VQC with a quantum gradient descent performs slightly better than the one with the classical descent, always in the high regimes.

In Ref.~\cite{2021Wu}, a VQC is applied to the analysis of Higgs boson production in association with a top-quark pair and a Higgs boson decay to two muons.
The goal is to train a supervised learning model able to discriminate between two processes, the signal events, $H\rightarrow\mu^{+}\mu^{-}$ and the background events; namely, a $Z/\gamma^{*}\rightarrow\mu^{+}\mu^{-}$ interaction.
A VQC-type model is proposed and trained based on thirteen kinematic variables associated with the process of interest.
A pre-processing step to reduce the feature space dimensionality is employed using the Principal Component Analysis (PCA) method.
A ten-qubit quantum circuit is set up using the variational circuit design in Ref.~\cite{Havlek2019}.
Only half of the qubits are measured to reduce the potential errors associated with measurement.
The circuit is trained through the noisy simulation of a ten-qubit system using the IBM's \emph{qasm$\_$simulator}.
An error mitigation scheme is applied to correct measurement errors through a relation matrix between the ideal and noisy results, and applied to the noisy results. 
The loss function is defined by the error probability of incorrect assignment compared to the exact solutions available for the training dataset.
During the training, the loss function is minimized to penalize missasignment and to optimize variational parameters using the SPSA optimizer.
The VQC performance is benchmarked against classical ML models, including a Support Vector Machine (SVM) constructed using the \emph{scikit-learn} package and a BDT set up using the \emph{XGBoost package} \cite{Chen:2016:XST:2939672.2939785}.
A hyperparameter tuning was performed on the classical models. The authors report an agreement between the results obtained when training on hardware and noisy simulation setups when error mitigation techniques are applied.
Furthermore, a similar performance is observed on models trained on simulated noisy settings, combined with error mitigation techniques and classical ML models.

In Ref.~\cite{Gianelle:2022unu}, a VQC approach has been used to identify the flavour of jets produced in proton-proton collisions at the LHCb experiment.
The identification of jets plays an important role in physics at hadron colliders, and the ability to distinguish the flavour of the quark generating a jet is fundamental in order to search for New Physics processes and find possible deviations from the Standard Model.
In this paper, jets initiated by $b$ or $\Bar{b}$ quarks are identified using a VQC, and its performance is compared with an exclusive algorithm used so far at LHCb, the so-called \emph{muon tagging} algorithm, which infers the jet flavour by means of the charge of the muon found inside the jet, and a standard Deep Neural Network (DNN).
LHCb simulated data are analysed, and jet substructure information is used as input features for the classifiers; for each jet, five types of particles (muon, kaon, electron, pion, and proton) with the greatest $p_T$ inside the jet are chosen.
For each type of particle, three variables are considered: the transverse momentum with respect to the jet axis ($p_{\mathrm{T}}^{\mathrm{rel}}$), the distance in the pseudorapidity-azimuthal plane to the jet axis ($\Delta R$) and the charge of the particle ($q$); a global variable, the weighted charge of the jet ($Q$), is also considered, accounting for a total of $16$ variables.
Two circuit geometries have been studied: the Angle Embedding geometry, where $n$ input features are mapped to angles $\theta$ of rotational gates applied to $n$ different qubits, thus resulting in a $n$-qubit circuit, and the Amplitude Embedding geometry, where input features are embedded in the amplitudes of a state vector of a $\lceil\log_2{n}\rceil$-qubit state.
The complexity of the circuit depends on the number of \emph{strongly entangling} layers, consisting of general rotational gates and CNOT \emph{entangling} gates, chosen in the variational part of the circuit.
In order to understand the impact of the considered variables in the classification, two versions of the dataset have been used: the \emph{muon dataset}, where only variables coming from the muon inside the jet plus the jet charge $Q$ are used, and the \emph{complete dataset}, which uses all variables.
In this way, for the \emph{muon dataset} the Angle Embedding (Amplitude Embedding) geometry is a circuit of $4$ ($2$) qubits, while for the \emph{complete dataset} the Angle Embedding (Amplitude Embedding) geometry uses $16$ ($4$) qubits.
Results show that the Angle Embedding geometry works better than the Amplitude Embedding geometry, reaching the same performance of the DNN for the \emph{muon dataset} while for the \emph{complete dataset} the Angle Embedding classifier has slightly worse performance than the DNN.
The study also shows that by increasing the number of strongly entangling layers, the classification accuracy increases up to a certain value where there is no further improvement, giving interesting suggestions on the optimal number of strongly entangling layers. Analysis of the number of training events shows that the quantum classifiers perform better than the DNN when using fewer training events while reaching similar performance when increasing the number of training events.
Finally, the impact of noise on the circuit geometries has been considered using simulations that account for circuit noise contribution: no evident degradation in performance is found, suggesting that these circuits are rather robust and, in principle, can run on hardware.

\subsubsection{Support vector machine}
\label{sec:svm}

Quantum Support Vector Machines (QSVMs)~\cite{Rebentrost2014} share many similarities with VQCs.
Both systems encode the classical input states in a Hilbert space by applying unitary designs tailored to the application.
While this encoding is conceptually equivalent in both approaches, the two models differ in how the quantum state is handled once it is prepared.
VQCs can be formulated as quantum kernel methods~\cite{SchuldKernel}.

The SVM algorithm maps $\vec{x}$ into a higher dimensional feature space, where it measures the similarity between any two data instances, denoted as ``kernel entries,'' $k(\vec{x_{i}}, \vec{x_{j}})$.
The SVM algorithm then optimizes a hyper-plane that separates data points into two categories.
A main limitation of the classical SVM algorithm is that evaluating kernel entries in a large feature space can be computationally expensive.
Thus, the QSVM is expected to leverage the quantum state space as a direct representation of the feature space, giving rise to  kernel functions that are hard to evaluate classically.
From this foundation, quantum feature maps can be designed and tested on practical datasets, potentially leading to better classification results than classical feature maps and kernels.
Once a suitable quantum feature map is chosen, the kernel matrix element is constructed by sampling the probability of measuring $|0\rangle$: 
\begin{equation}
    k(\vec{x_{i}}, \vec{x_{j}}) = |\langle 0^{\otimes n}|U^{\dagger}(\vec{x_{i}})U(\vec{x_{j}})|0^{\otimes n} \rangle|^{2}.
\end{equation}
Another main difference with VQCs is that the loss function of a QSVM depends on the inner product of the feature vectors, \textit{i.e.}, the goal is to maximize
\begin{equation}
    L(c_{1}...c_{n}) = \sum_{i=1}^{n}c_{i} - \frac{1}{2}\sum_{i=1}^{n}\sum_{j=1}^{n}y_{i}c_{i}(\vec{x_{i}}\cdot\vec{x_{j}})y_{j}c_{j},
\end{equation}
subject to $\sum_{i=1}^{n}c_{i}y_{i}=0$, and $0\leq c_{i} \leq \frac{1}{2n\lambda} \equiv C$ for all $i$.
Here, $c_{i}$ are the independent variables of the loss, $\vec{x_{i}}$ and $\vec{x_{j}}$ are feature vectors of a given pair of data points, $i$ and $j$ and $y_{i},y_{j}$ their corresponding labels.
Finally, $n$ is the number of encoded features, and $\lambda$ is a regularization parameter that tunes the trade-off between mis-classification and width of the SVM margin. 

Furthermore, in QSVM-like circuit designs, all quantum registers need to be measured to construct the quantum kernel, whereas in VQCs this is not a requirement.

The Higgs classification problem of Ref.~\cite{Belis2021} also studied a classification scheme based on the implementation of a QSVM.
Again, auto-encoders are used to reduce the dimension of the feature space, as explained in Sec.~\ref{sec:vqc}.
In terms of feature maps, two data embedding circuits are tested: 1. An amplitude encoding circuit with N qubits, capable of encoding $2^N$ features, for a 4-qubit and 6-qubit architecture; 2. An 8-qubit architecture, which is more suitable for an implementation on a NISQ device.
The 4- and 8-qubit models have a similar performance to a classical SVM with a radial base function kernel.
Likewise, the 6-qubit QSVM with amplitude encoding performs similar to an SVM with a linear kernel.

In Ref.~\cite{2021WuSunGuan}, a QSVM model is used again for the discrimination of $t\overline{t}H$ events in the $H\rightarrow \gamma\gamma$ decay channel from non-resonant two-photon production events.
Three classical kernels are considered to benchmark the classical SVM method: the linear kernel, the polynomial kernel, and the radial basis function (RBF) kernel.
Both quantum and classical SVM models are trained using 15 kinematic variables resulting from the compression of the original 23 kinematic variables available in the dataset, by using a PCA method.
The 20-qubit model is trained on a noise-less simulator and showed similar performance compared to the classical SVM. 
When deployed on hardware, the system size is reduced to 15 qubits, and the authors report similar performance compared to the noiseless simulation. 

Another implementation of a QSVM algorithm for classification is described in Ref.~\cite{Heredge2021}.
Here, the authors designed and implemented a QSVM approach for the signal-background classification task in $B$ meson decays.
Three different encodings were tested to study the impact of the choice of feature map in the classification performance.
In particular, the Bloch encoding circuit, which the authors designed for encoding particle data, outperformed the other two and used fewer resources for the same task.
When using a limited number of inputs, the QSVM outperformed classical methods in simulations.  

In Ref.~\cite{Peters2021}, an extension to the method of machine learning based on quantum kernel methods is extended up to 17 hardware qubits requiring only nearest-neighbor connectivity.
This circuit structure is used to prepare a kernel matrix for a classical SVM to learn patterns in 67-dimensional supernova data for which competitive classical classifiers fail to achieve 100$\%$ accuracy.
Furthermore, the circuit design is justified based on its ability to produce large kernel magnitudes that can be sampled to high-statistical certainty with relatively short experimental runs.
This experiment is, by far, the largest classification model deployed on hardware and demonstrates similar performance to non-quantum techniques for the classification of a HEP dataset.

\subsubsection{Quantum Convolutional Neural Networks}
\label{sec:qcnn}
In Section~\ref{sec:vqc}, we introduced the concept of VQCs and compared the tunable parameters to the weights in a classical neural network.
We now introduce the concept of quantum convolutional neural network (QCNN).
QCNN is the framework that uses VQCs to perform the convolutional operations in a classical CNN.
Convolutional filters or kernels are replaced with VQCs to harvest the expressive power granted by quantum entanglements.
The quantum convolutional kernels will sweep through the input data and transform them into a representation vector of lower dimensions by performing measurements.
A stack of VQCs will ensure features of varied length scales that are captured in different layers. 
In Ref.~\cite{chen2020quantum}, a QCNN framework for the classification of HEP events from the simulated data from the DUNE experiment is proposed.
Based on the success of classical CNNs for the classification of images, a QCNN is proposed for the classification of neutrino events as detected in a Liquid Argon Time Projection Chamber (LArTPC).
This technology allows for detecting neutrino events through high-resolution images of particle interactions within the detector volume as the ionized electrons drift towards the multiple sensing wire planes.
The goal of the QCNN is to predict the types of different particles by analogy with those performed via classical CNN.
The target labels correspond to the four possible particle types to be detected, an electron, a muon, a pion, or a proton.
The authors report that the quantum model can learn faster and reach better testing accuracy with fewer training epochs, when using a similar number of parameters in the classical and quantum CNN benchmarks on noiseless simulations.

In Ref.~\cite{chen2021hybrid}, a QCNN based technique aimed at understanding how QML models can provide advantages over classical models when dealing with sparse data, which is common in scientific data.
This work introduces a hybrid quantum-classical graph convolutional neural network (QGCNN) framework applied to the same classification problem in Ref.~\cite{chen2020quantum}.
This study compares the performance of a QGCNN to a classical multilayer perceptron and CNNs and reports that the 10-qubit quantum model requires fewer parameters to achieve comparable performance as the classical models.
Furthermore, they compare the QGCNN model to the QCNN model proposed in Ref.~\cite{chen2020quantum}.
Both quantum models can achieve similar performance, but the QGCNN requires half the number of parameters used in the QCNN, highlighting the importance of models tailored for sparse datasets.

\subsubsection{Anomaly detection models} 
\label{sec:qvae_classification}
The search for new physics depends on our ability to separate Standard Model events from the background's much rarer and complex signal events.
A new approach for finding events consistent with beyond-the-standard-model physics is through data-driven search, where no assumptions on the new physics scenario are made.
While this allows for a search on a broader phase space, it requires a classification technique that can successfully learn and identify the important features in the background data to discriminate them from rare signal events which do not share the same properties.
In classical ML, autoencoder architectures have been designed for this purpose~\cite{vaes_classical}.
Autoencoder architectures consist of an encoder step that compresses the feature space into a latent space with reduced dimensionality.
Subsequently, the latent space is decoded into an output of the same dimensionality as the input feature space.
The entire network is then trained such that the loss function, which evaluates how well the output resembles the input, is minimized.
Because quantum mechanics can generate patterns with properties beyond classical physics, a quantum computer should be able to recognize patterns beyond the capabilities of classical machine learning.
Thus, the motivation for a quantum autoencoder is that such a model would allow us to efficiently perform the dimension reduction of quantum data.
In classical architectures, the necessary compression and expansion of data in the encoding and decoding steps are manifestly non-unitary, which has to be addressed by the QAE using entanglement operations and reference states which disallow information to flow from the encoder to the decoder.
A model capable of effectively reducing the dimension of a quantum state was initially proposed in Ref.~\cite{Romero2017}.
In the context of anomaly detection, the assumption is that the minimal dimension of the latent space for which the input features can still be reconstructed corresponds to the dimensionality of the information space required to describe the training sample, here, the SM background processes.
If the signal is kinematically sufficiently different from the background samples, the loss function or reconstruction error will be larger for signal than for background events.
In Ref.~\cite{2021BlancePhotonic}, an alternative approach based on continuous-variable quantum computing is proposed.
The authors use gaussian boson sampling to embed classical data into a feature vector and use an autoencoder scheme for model-independent searches through anomaly detection techniques.
A continuous variable alternative to the classical $k$-means clustering algorithm is devised and denominated Q-means clustering, potentially scaling to large feature vectors more efficiently.
The Q-means algorithm has a complexity in the order of $\log(N)$, with respect to the size $N$ of the feature vector.
This represents an improvement from the classical $k$-means algorithm with a complexity in the order of $N$.
The model is yet to be deployed in a continuous variable architecture, and the authors limit their implementation to discrete qubit quantum computing.
In Ref.~\cite{2021Ngairangbam}, the authors study quantum autoencoders based on VQCs for the problem of anomaly detection at the LHC.
For a QCD $t\bar{t}$ background and a resonant heavy Higgs signal, they find that a simple quantum autoencoder outperforms dense classical autoencoders for the same input space. 
In Sec.~\ref{sec:quantum_gen_mods}, QVAEs are discussed in the context of generative tasks, although no application in HEP has been reported.

\section{Data Generation/Augmentation}
\label{sec:data_generation}
A crucial element of any analysis in HEP involves the simulation of the physical processes and interactions taking place at HEP facilities to develop new theories and models to explain experimental data and characterize background, study detector response, and plan for detector upgrades.
The corresponding simulations start with a Lagrangian, compute the hard scattering in perturbative QCD, the parton shower in resummed QCD, model hadronization based on reliable precision measurements, and finally, a full detector simulation.
Technically, all of the processes involved rely on Monte Carlo techniques. 
These simulations are often computationally intensive, taking up a significant fraction of the computational resources available to physicists.
The production of reliable and statistically significant samples represents a huge computational footprint both in terms of CPU usage and disk space.
Currently, the best estimates suggest that the simulation of a single collider event already takes several minutes~\cite{Albrecht_Roadmap}, with $\bigO{109}$ events to be generated (on average) for each simulation campaign in the LHC. 

Generative models are trained to prepare a target distribution that can accordingly be used to generate new samples.
This is more challenging than the discriminative tasks described in Sec.~\ref{sec:classification} since it requires one to efficiently learn, represent, and sample from high-dimensional probability distributions, but quantum processors are well-suited for this task.  
Many kinds of generative models used in classical machine learning use neural networks and are parameterized by weights and biases (\textit{e.g.} Boltzmann Machines (BMs)~\cite{ACKLEY1985147} and RBMs ~\cite{smolensky_1986}, autoencoders and variational autoencoders ((V)AEs)~\cite{kingma2013auto} and Generative Adversarial Networks (GANs)~\cite{GAN}).
Training generative models can be based on minimizing the energy of the model~\cite{ACKLEY1985147,fischer2014training}, minimizing the error when sampling from a target posterior and the model posterior~\cite{kingma2013auto}, or through adversarial methods~\cite{GAN}.
These neural network models have been translated into quantum models (q(R)BMs, q(V)AEs, qGANs) either as standalone VQCs (see Section \ref{sec:vqc}) or as a component in a hybrid network \cite{Romero2017}.  Many of the training workflows can be adapted as variational algorithms. The training leverages classical methods and hybrid workflows to optimize the parameterized quantum models.

The ability of quantum information processors to represent vectors in $N$-dimensional spaces using $\log{(N)}$ qubits, and to perform manipulations of sparse and low-rank matrices in time $\bigO{\mathrm{poly}(\log(N))}$~\cite{Lloyd2018} motivates the exploration of quantum generative models as an alternative to classical generative models for generative tasks in HEP.
Quantum generative models are expected to exhibit an advantage over classical generative models in runtime and the number of parameters needed to learn data distributions due to their strong expressive power.
Generative modeling has been used as near-term applications for quantum processors \cite{Benedetti2019ddqcl}, but as of yet not all models have been used in HEP applications. In the remainder of this section, we will overview the state-of-the-art results for two of the models above -- the q(R)BM and the q(V)AE. In separate follow-on sections, we provide an in-depth discussion of two quantum generative models, the qGAN and the quantum circuit Born machine (QCBM), that have been used in HEP applications.
\subsection{Quantum Generative Models}\label{sec:generative_models}
\label{sec:quantum_gen_mods}
\begin{figure}
    \centering
    \includegraphics[width=.98\linewidth]{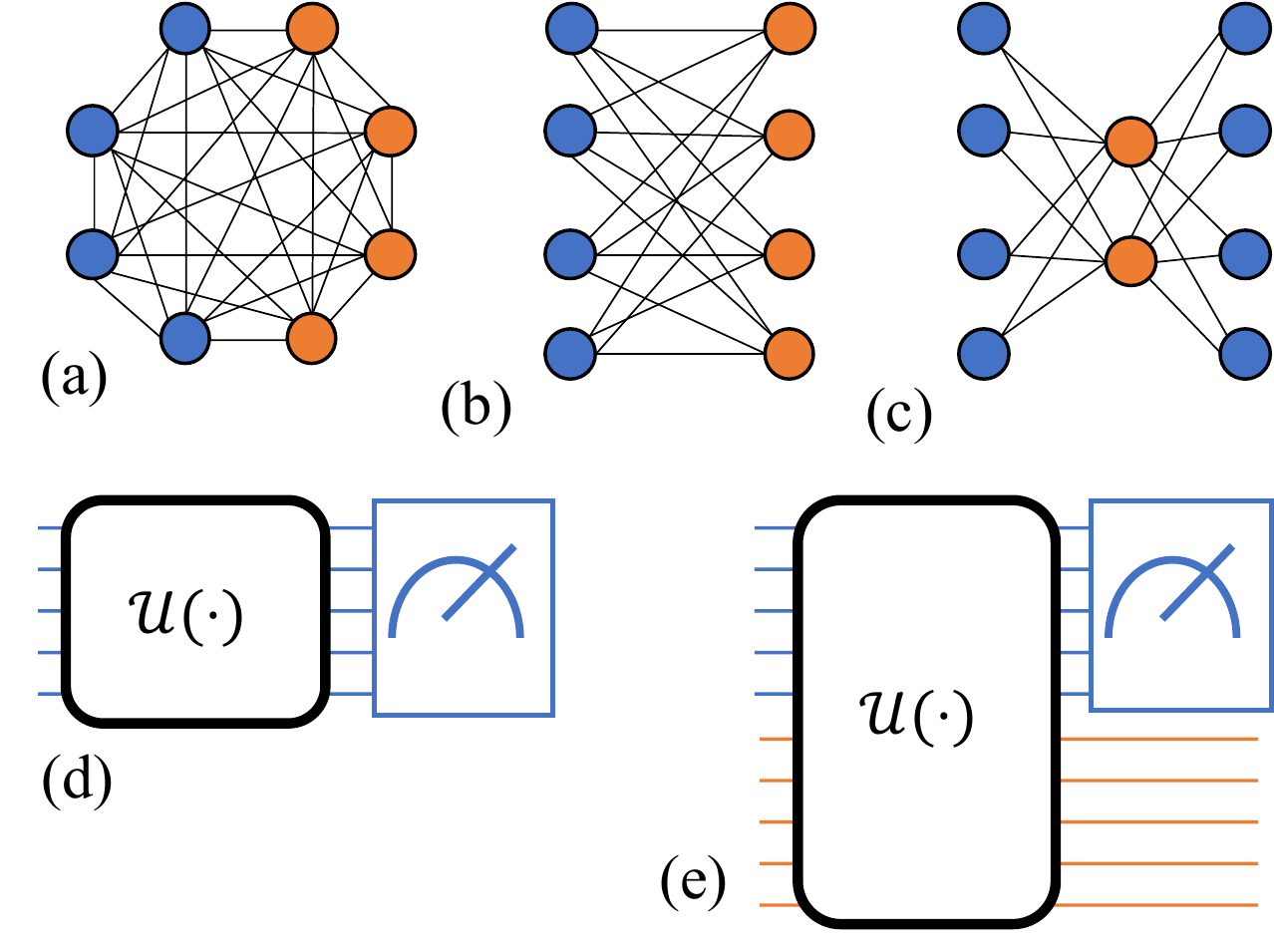}
    \caption{Classical generative models using neural network constructions with visible (blue) and hidden (orange) units (top row): \textbf{(a):} Boltzmann machine, \textbf{(b):} Restricted Boltzmann machine, \textbf{(c):(Variational)} Autoencoder.
    Quantum generative models using parameterized unitaries $\mathcal(U)(\cdot)$: \textbf{(d)}: Parameterized unitary circuit \textbf{(e)}: Parameterized unitary circuit with partial measurements.}
    \label{fig:generative_models}
\end{figure}
The (R)BM is a physically motivated neural network capable of generating new samples similar to the training data~\cite{ACKLEY1985147}, which was previously introduced in Section \ref{sec:classification}. 
A network of classical spins are connected by weighted edges ($w_{ij}$) and on-site biases ($b_i$), which are optimized during training to find the ground state of the system's Hamiltonian
\begin{equation}
    H_{(R)BM} = \sum_i b_i z_i - \sum_{\langle i, j\rangle} w_{ij}z_i z_j.
\end{equation}
BMs a natural fit for quantum annealing~\cite{PhysRevA.94.022308, Adachi_RBM, amin2018quantum} and quantum computing~\cite{Wiebe_DL, PhysRevX.8.021050,Zoufal2021}; in the latter Ref.~\cite{Zoufal2021} shows polynomial speed-ups relative to classical training. 
Starting from a classical (R)BM, the neurons (or classical spins) are replaced by qubits and the Hamiltonian is translated into a unitary circuit or Ising model couplings.  

VAEs~\cite{kingma2013auto} are artificial neural networks typically used for generative modeling applications.
The word \textit{autoencoder} refers to the fact that its architecture, comprising an encoder and a decoder, resembles a traditional autoencoder.
Nonetheless, there are significant differences, especially in the mathematical formulation.

The goal of an autoencoder~\cite{kramer1991nonlinear} is learning to map some input vector $x$ to a compressed space vector $z$.
In the language of generative modeling, this compressed space is known as the \textit{latent space}.
This latent space vector is fed into another network that reconstructs the input.
The network that compresses the input vector $x$ is known as the encoder, whereas the network that maps the latent vector $z$ to the initial input $x$ is the decoder.
A figure of merit for the loss function training is simply the error of the reconstructed input state, which shall be minimized.

Notice, however, that there are no constraints on the latent variables $z$.
This is the main reason why traditional autoencoders are useful for compressing information but lack a generative structure. Namely, the latent variables $z$ rarely capture relevant information, or this information is not organized. This is the precise issue that VAEs tackle.
To be specific, VAEs deal with the problem of the latent space irregularities by making the encoder return a simple distribution over the latent space, generally a Gaussian distribution, by adding extra terms in the loss functions to ensure a better organization of the latent space~\cite{bengio2013representation}.

The encoder and decoder of a qVAE model are gates which transform an underlying wavefunction. At any point in the network, the wavefunction defines a distribution over a set of basis states meaning that non-Gaussian distributions can be prepared in the latent space. In Section it was discussed how qAEs have been used for anomaly detection \ref{sec:classification}.  Other applications for qAEs include quantum data compression~\cite{Romero2017} and qVAEs have been used for generative modeling of classical data.
In Ref.~\cite{khoshaman2018quantum} a quantum generative process is used to produce handwritten digits, with competitive results on par with other
classical methods. 

\subsection{Quantum Generative Models for HEP}
\label{sec:qGANs}
For HEP applications, two quantum generative models have been used: the qGAN \cite{Lloyd18_QGAN,QGAN_Zoufal19} and the QCBM \cite{Zhao18_QCBM,Liu18_QCBM}. There are many GAN architectures. However, all of them contain as basic components a generator, a discriminator, and an adversarial training procedure as suggested in the initial work of Ref.~\cite{goodfellow2014generative}. The qGAN follows the general structure of a classical GAN, whereas the QCBM uses only a parameterized circuit. 

In general the generator $G$ transforms samples from some prior, simple-to-sample noise distribution $z \sim p_{\mathrm{prior}}(z)$ into new samples $G(\phi_g)$, thus mapping $p_{\mathrm{prior}}(z)$ to a different distribution $p_{\mathrm{fake}}$ of generated data.
The discriminator takes as input samples $x$, from the real distribution in the form of data, and tries to distinguish between fake data from the generator and this real data $p_{\mathrm{real}}$.
The adversarial training procedure alternates between the generator suggesting samples that are accepted by the discriminator and the discriminator detecting those samples suggested by the generator.
Once this adversarial game reaches a stable equilibrium over training epochs, the generator robustly suggests new, previously unseen, samples of the target distribution.

In practice, often the binary cross-entropy is used as optimization objective.
The individual loss functions can then be defined as
\begin{align}
   \mathcal{L}_G(\phi_g,\phi_d) &= -\mathbb{E}_{z \sim p_{\mathrm{prior}}(z)}[\log D(\phi_d,G(\phi_g,z))]  \,,\\\nonumber
   \mathcal{L}_D(\phi_g,\phi_d) &= \mathbb{E}_{x \sim p_{\mathrm{real}}(x)}[\log D(\phi_d,x)] \\ &+\, \mathbb{E}_{z \sim p_{\mathrm{prior}}(z)}[\log (1-D(\phi_d,G(\phi_g,z)))]\,,
\end{align}
for the generator and discriminator, respectively. As such the training corresponds to a minimax two-player game,
\begin{align}
 \underset{\phi_g}{\min}\,\,\mathcal{L}_G(\phi_g,\phi_d)  \,,~~~
 \underset{\phi_d}{\max}\,\,\mathcal{L}_D(\phi_g,\phi_d)  \,,
\end{align}
where the optimum uniquely corresponds to the Nash equilibrium between the loss functions.

The neural network architecture, input data, and complexity of the distributions all play a role in successfully training a GAN, both in the classical and the quantum context.
Here, in particular, the potential to create more complex, varied outcomes from simpler network architectures in the quantum case could prove valuable.
Studying this connection between quantum network architecture and complex outputs in detail provides interesting future opportunities.
For example, in Ref.~\cite{Liu2021} a robust training advantage was established in the context of supervised learning.

In HEP data analysis, two applications stand out for their intensive computing resource requirements.
These are the simulation of the detector and its various calorimeters on the one hand and the generation of process events via Monte-Carlo on the other.
Both fields are under intensive research using classical methods due to the potential gains from any speed-up.
Finding a path towards quantum advantage and speed-up could lead to a jump in progress in the future.

\subsubsection{Detector simulations}
The traditional Monte-Carlo-based simulations of HEP detectors are highly time-consuming.
As an example, it is estimated that during the last years before the last LHC shutdown, the LHC experiments devoted more than 50 \% of their  WLCG computing resources to Monte-Carlo production~\cite{hsf_white_paper}.
At the HL-LHC, given the higher detector granularity, the larger complexity of physics events, and the statistics needed for accurate precision measurement, Monte-Carlo production is expected to increase by at least two orders of magnitude~\cite{hsf_white_paper}.
In this context, deep generative models and, in particular,  Generative Adversarial Networks, have been investigated in a range of different use cases as possible alternatives for fast simulations.
Examples include 3DGAN~\cite{Vallecorsa2018, GulRuk2018} or CaloGAN~\cite{Paganini2018}. 

In most cases, the GAN-based prototypes reach a high level of accuracy and produce realistic datasets.
Given the increasing interest in quantum generative modeling, it is only natural to explore the possibility of replacing classical generative models using quantum or hybrid classical-quantum architectures. 
Example applications exist, proposing hybrid quantum GAN, where a quantum generator is trained against a classical discriminator to reproduce the output of electromagnetic calorimeters~\cite{Zoufal2021}.
The Dual-Parameterized Quantum Circuit (PQC) GAN, is characterized by a classical discriminator and two quantum generators that take the form of PQCs, and it is capable of simulating reduced size pixelated images of calorimeter volumes~\cite{DualPQC}.
The model was trained to reproduce one-dimensional energy distribution, generated inside the calorimeter volume by single electrons.
The studies in Ref.~\cite{ACAT_2021} show that the results are stable on both simulated and real quantum hardware.  
Similarly, the work in Ref.~\cite{CVqGAN} explores the performance of a hybrid prototype (based again on a quantum generator and a classical discriminator) implemented using photonic-based quantum computing~\cite{CVNNs}, which has the great advantage of allowing natural representation of continuous variables in quantum states.

Despite the encouraging initial results described above, it should be noted that the use case of detector simulation presents important challenges related to the size of the distributions the generative models learn.
In most cases, this scales with the number of sensors in the detectors and it is therefore challenging to embed in quantum circuits, especially on near-term devices. 
 
\subsubsection{Monte Carlo event generation}

The use of GANs for applications in the simulation of HEP events has been a key focus in the context of classical machine learning~\cite{baldi2021gan,Backes_2021,butter2020generative,Butter_2021,Butter_2020,Bellagente_2020,Butter_2019}.
Due to their ability to generate larger samples of a distribution given through a more limited data set, they present a unique opportunity for data augmentation and signal boosting applications. 


An example application is to provide distributions of key experimental processes and to generate a larger sample using a qGAN.
This was successfully implemented for the process of $pp\rightarrow t\bar{t}$ production at the LHC with $\sqrt{s} = 13$ TeV in~\cite{bravoprieto2021stylebased}.
Based on the data re-uploading idea~\cite{perezsalinas:2020reuploading}, there a flexible network architecture was proposed that enabled an efficient representation of the desired generative model. 

In particular, through hyperparameter tuning, it was found that a network in which each input dimension of the distribution is represented by a single qubit, entangled with its neighbors, and a single layer was sufficient to successfully train the GAN with a robust minmax condition found (with a few tens of thousands of epochs of training, comparable with classical experiences).

The low complexity of the quantum circuit used enabled its deployment on both superconducting and ion-based quantum computing architectures.
Good signal-to-noise properties were observed in both cases. 

Even though the work in this direction is at a proof-of-principle stage, results like these give hope for successful deployment on NISQ devices.

Further refinements, such as training efficiently for other processes through transfer learning, the inclusion of error correction techniques in the generation process, or further implementation of quantum components - as e.g. the given example implements a classical discriminator - are useful avenues to follow in the future. 

The use of the trained qGAN is not limited to data augmentation purposes.
Additionally, it can also be used as an intermediary encoder of the process generating functions.
In Ref.~\cite{Agliardi:2022ghn} it was successfully demonstrated that a qGAN could be used to load the underlying information of the process onto qubits.
A quantum amplitude estimation (QAE) algorithm subsequently performed the integration of the elementary process in one and two dimensions on these loaded qubits. 
In future, it would be interesting to pursue further opportunities to combine different quantum algorithms in this way.
One could envision the construction of a ``quantum pipeline'' from raw data to refined results for physics observables in which different quantum algorithms are connected in such a way that they benefit from practical quantum advantages.
For example, one could imagine feeding in data from an efficiently trained and sampling quantum network~\cite{Huang:2021ydj,huang2021information} into subsequent quantum analysis tools, such as the QAE, via qGANs or exact encodings.

The QCBM is an example of an implicit model for generative learning~\cite{MohamedLearning} that generates data by measuring the system as a Born machine \cite{Benedetti2019ddqcl,coyle2020born}.  Unlike the q(R)BM models, the unitary circuit is not defined with respect to a specific Hamiltonian, rather a QCBM is a parameterized unitary $U(\vev{\theta})$ that prepares $N$-qubits in the state $|\Psi_{\Theta}\rangle = U(\vec{\theta})|\Psi_{0}\rangle$. One can obtain a classical distribution over the $2^{N}$ computational basis states by measuring 
$|\Psi_{\Theta}\rangle$ in a fixed basis (usually the computational basis). Unlike the qGAN model, QCBMs can be trained using non-adversarial methods, either using gradient-free \cite{Benedetti2019ddqcl} or gradient-based \cite{PhysRevA.98.062324} optimization. Training a QCBM is done by minimizing a loss function $\mathcal{L}(P_{target},P_{QCBM})$ that computes the similarity between the target distribution and the distribution sampled from the QCBM, such as the Kullback-Leibler (KL) divergence, maximum mean discrepancy (MMD), Stein discrepancy, Sinkhorn divergence or optimal transport \cite{coyle2020born}. The study in Ref.~\cite{DelgadoHamiltonQCBM}, uses non-adversarial gradient-based training, more specifically the KL divergence, of 8- and 12-qubit QCBMs to generate joint distributions over 2 and 3 variables to generate synthetic data of a typical HEP process. In Ref. \cite{QCBM_Kiss22}, a conditional QCBM is introduced and trained with the MMD loss to generate (multivariate) Monte Carlo event as a function of the incoming energy.


\section{Quantum-inspired Algorithms}
\label{sec:quant-insp}
The fusion of quantum computing and data analytics promises a revolution in machine learning and other optimization and related computational capabilities.
However, there are only a few demonstrations of quantum advantage for very specific tasks.
Given the current hardware and software limitations, an exciting exploration for HEP scientists is quantum-inspired algorithms, viewing information technologies from the vantage point of quantum to make gains in conventional systems. 

Quantum-inspired algorithms involve a traditional (non-quantum) computer emulating certain aspects of quantum mechanics to gain a computational advantage.
To date, there have been quite a few quantum-inspired algorithms tackling problems of optimization~\cite{CAI2021114629}, machine learning~\cite{Ding2021}, and linear algebra~\cite{Chakhmakhchyan2017}.
In addition, the Fujitsu Digital Annealer~\cite{Aramon2019} is a classical analog of quantum annealers such as those produced by D-Wave.

In HEP, a method of increasing popularity rooted in quantum mechanical concepts is tensor networks (TNs)~\cite{Bridgeman_2017, Biamonte_TNinAShell} or tensor network states.
TNs have been developed to investigate quantum many-body systems on classical computers by efficiently representing the exponentially large quantum wavefunction in a compact form. 

Recently, it has been shown that TN methods can also be applied to solve ML tasks very effectively~\cite{Stoudenmire2018, Levine2017, NIPS2016_5314b967}, yielding comparable results when benchmarked against NNs on standard datasets~\cite{Stoudenmire2018, NIPS2016_5314b967, Glasser2018}.
In HEP, TNs have been explored in an ML context to gain insight into the learned data by computing quantum correlations or entangling entropy.

In Ref.~\cite{Felser2021}, a TN-based supervised learning approach to the identification of the charge of $b$ quarks (\textit{i.e.}, $b$ or $\bar{b}$) is presented. Simulated data of $b$ quarks generating jets produced in high-energy proton-proton collisions at the LHCb experiment are studied.
Although the TN-based method yielded similar performance compared to a deep neural network classifier, some benefits of the TN-based approach were highlighted, including the power of compressing the network while keeping a high amount of information.

Similar conclusions were reported in Ref.~\cite{Araz2021}, where a TN-inspired classifier is used to discriminate top jets over QCD jets.
It is also noted that the TN learns the volume and correlations of the projected geometry of topological relations in the data, which is reflected by the entanglement entropy of the network.
This observation can be exploited to reduce redundant information in the input data, thereby reducing the complexity of the network while maintaining a high classification performance.

\section{Challenges}
\label{sec:challenges}
In this section, we describe some of the limitations that are common to the applications discussed in this manuscript. 

\subsection{Quantum annealing-based algorithms}


Compared with other quantum computing paradigms, quantum annealing has the advantage that there are already commercially available processors with thousands of qubits, and these are expected to grow further in the near term.
Nevertheless, is remains elusive a clear experimental demonstration of quantum advantage for quantum annealing.

Moreover, quantum annealers are severely limited in the tasks that they can perform. 
Specifically, they are only designed to solve quadratic unconstrained binary optimization (QUBO) problems \cite{arthur2021balanced, date2021adiabatic}, although the more general quantum adiabatic computing model is universal \cite{AharonovAQC}. 

A common problem with annealing-based solutions for HEP data analysis is the large overhead of converting the problem at hand to the corresponding QUBO formulation, that is, computing all the QUBO coefficients.
The time taken in this pre-processing step largely differs from problem to problem, but, as we have seen, often scales worse than quadratically with instance size.
Another challenge is the required annealing schedule to reach the global minimum, which determines the final complexity of the algorithm.
This is difficult to estimate for real-world data analysis problems, but for many applications it is expected to scale exponentially with the number of variables in the problem.

Additionally, if a problem requires a significant degree of connectivity among its binary variables (also called logical qubits), it is extremely difficult to embed it onto the physical qubits.
Indeed, some instances might take exponentially long for the embedding process \cite{date2019efficiently}.

For HEP problems specifically, the current methods for problem decomposition such as \texttt{qbsolv} by D-Wave are inadequate because of their lack of precision required for collider experiments.
These problems also require significant hyperparameter optimization in addition to the pre- and post-processing subroutines for dealing with noise, freeze out, and competing strengths between spin couplings in the Hamiltonian and the chain strengths.

\subsection{Gate-based/universal QC algorithms}

Gate-based quantum computers have already been used to establish a clear advantage over the classical counterparts \cite{Arute2019}.
However, present-day universal quantum computers are noisy and limited in the number of qubits to serve many practical applications, in particular in HEP data analysis.

In the HEP context, a significant challenge is the interface between the classical data and the quantum computing architecture.
Some proposals assume access to QRAM~\cite{QRAM}, which would allowed coherent access to previously stored classical data.
However, despite existing theoretical proposals for building such devices~\cite{QRAM_architectures}, it is not expected that they could be implemented in the near term.

Considering solutions that do not require fault-tolerance, the current generation of quantum variational circuits often uses the amplitude or basis encoding schemes, as mentioned in Sec.~\ref{sec:vqc}.
While these are important first steps, each scheme has some shortcomings.
For instance, the data that can be stored using the basis embedding protocol is linear in the number of qubits being used, much smaller than the Hilbert space, which grows like $2^{N}$ \cite{date2020quantum}.
On the other hand, the amplitude encoding scheme does allow for storage of $2^{N}$ variables but is less effective in certain contexts~\cite{Gianelle:2022unu}.
As has been pointed out in~\cite{PhysRevLett.122.040504}, this may be due to the fact that non-linear transformations of the data are difficult to design in this context.
However, how to move beyond these schemes is not obvious as other encoding mechanisms require a large amount of preprocessing on the input data; see \textit{e.g.} Ref.~\cite{jlazar2021}.
Going forward, we will need to find methods for efficiently and densely embedding data in quantum registers in order to take full advantage of QC's expressiveness.

In addition to the inefficiency in embedding data onto the quantum computer, there is a lack of knowledge on the kind of problems where one embedding scheme might be better suited than the other.
Furthermore, the understanding about how noise might impact the performance of training is somewhat unclear---while some studies advocate for noise adding much desired stochasticity during training, other studies present noise as a hindrance.
In terms of logistics, accessing quantum computers on the cloud remotely and running jobs on them is extremely inefficient---in some cases, users have to wait for days before their jobs are run on the hardware.

Finally, it remains an open challenge to find a quantum algorithm for HEP data analysis with proven \emph{exponential} speedup.
In general, such algorithms are rare~\cite{shorperspective}, requiring highly structured problems~\cite{Aaronson2014}.
Importantly, Ref.~\cite{Babbush2020} argues that with the current error-correction techniques we should not see a time advantage for problems with small polynomial speedups, at least when we consider practical input sizes.

\section{Outlook}
\label{sec:outlook}
Over the last couple of years, quantum computing techniques have been developed to explore the applicability of quantum computing to data analysis in HEP, promising better algorithms for ML, optimization, and other techniques.
In this manuscript, we highlighted some areas where quantum computing applications have been explored for data analysis in HEP.
The major drawback of these ideas is the lack of a fully mature quantum computer.
The main limitations were discussed in Section~\ref{sec:challenges}.
Naturally, the first question that arises from this statement is whether there will ever be a demonstration of quantum advantage. The second question to be answered is if there is a short-term gain in exploring quantum computing for data analysis in HEP.
While it may be sometime before quantum computers mature enough to replace classical computing in certain HEP analysis workflows, there are quantum explorations appropriate even for today’s data analysts.
These applications can be categorized into main areas: (1) how HEP can benefit from the use of QC, and (2) how HEP can contribute to the quantum computing ecosystem.

In Section~\ref{sec:quant-insp}, we discussed how quantum-inspired algorithms might provide a new perspective to the main computational problems facing the HEP community.
Due to the requirement of analyzing vast, highly correlated data in order to exploit the full physics potential of the LHC or other large sample size particle physics experiments such as DUNE or IceCube, it becomes more and more critical to develop a fundamental understanding of the data analysis methods applied.
In this context, TNs are able to extract information not easily accessible to NNs, such as correlation functions and entanglement entropy which can be used to explain the learning process and subsequent classifications, paving the way to an efficient and transparent ML tool.
Furthermore, TNs have been shown to reduce prediction times.
In Ref.~\cite{Felser2021}, it is expected that in the near future, prediction times will be reduced to the order of MHz, having the potential for deployment on FPGAs for real-time data acquisition and selection at HEP facilities.

As the size and quality of QCs continue to advance, with large-scale entanglement achieved on a range of platforms, so does the feasibility of using quantum machines to perform typical data analysis tasks in particle physics.
In particular, there are various ways in which the field of ML may benefit from the advent of QC.
These benefits range from speed-ups to specific subroutines, such as gradient descent and linear algebra, to quantum analogs of classical algorithms.

Finally, quantum computing benchmarks on HEP experimental data can be used to formulate metrics and benchmarks for emerging quantum computers and other quantum technologies.
For reasons of cross-platform comparison and the identification of trends to make predictions, metrics and benchmarks have an important role to play even today, when quantum computers cannot, with rare exception, outperform conventional devices.
One example is jet substructure, where analytics reconstruction techniques coexist with numerical MVA methods. 
The combination of a large amount of available data with an excellent theoretical understanding of the underlying physics in collider phenomenology provides an ideal environment to explore novel quantum and quantum/classical techniques.

\begin{acknowledgments}
This work was partially supported by the Quantum Information Science Enabled Discovery (QuantISED) for High Energy Physics program at ORNL under FWP ERKAP61. This work was partially supported by the Laboratory Directed Research and Development Program of Oak Ridge National Laboratory, managed by UT-Battelle, LLC, for the U. S. Department of Energy. 

DM, YO, and PB acknowledge  project \textit{QuantHEP – Quantum Computing Solutions for High-Energy Physics}, supported by the EU H2020 QuantERA ERA-NET Cofund in Quantum Technologies and FCT -- Funda\c{c}\~{a}o para a Ci\^{e}ncia e a Tecnologia (QuantERA/0001/2019), and thank the support from FCT, namely through project UIDB/50008/2020. 
DM acknowledges the support from FCT through scholarship 2020.04677.BD. AG, DL, DN, LS, and DZ acknowledge the support from the Department of Physics and Astronomy "Galileo Galilei" of University of Padova and from INFN (Istituto Nazionale di Fisica Nucleare).
\end{acknowledgments}

\bibliographystyle{unsrt}
\bibliography{bibliography}

\end{document}